\documentstyle[12pt]{article}

\def\gsim{\mathrel{%
   \rlap{\raise 0.511ex \hbox{$>$}}{\lower 0.511ex \hbox{$\sim$}}}}
\def\lsim{\mathrel{
   \rlap{\raise 0.511ex \hbox{$<$}}{\lower 0.511ex \hbox{$\sim$}}}}

\title{Primordial space-time foam as an origin of cosmological 
matter-antimatter asymmetry\thanks{This is an extended version of an essay
selected for an ``honorable mention'' in the Annual Essay Competition of 
the Gravity Research Foundation for the year 2001.}}

\author
{D.~V.~Ahluwalia$^{a,b}$, 
M.~Kirchbach$^a$\thanks{E-mail: ahluwalia@phases.reduaz.mx, 
kirchbach@chiral.reduaz.mx}\\ 
$^a$ISGBG, Facultad  de  Fisica de la UAZ,\\ 
Ap. Postal C-600, Zacatecas 98062, 
Mexico\\
$^b$Mail Stop H-846, 
Los Alamos National Laboratory\\ Los Alamos, NM 87545, USA}

\begin{document}

\def\beq{\begin{eqnarray}}
\def\eeq{\end{eqnarray}}



\maketitle


\begin{abstract}
The possibility is raised 
that the observed cosmological matter-antimatter asymmetry 
may reside in asymmetric space-time fluctuations and their interplay 
with the St\"ckelberg-Feynman interpretation of antimatter. The presented
thesis also suggests that the effect of space-time fluctuations is to diminish
the fine structure constant in the past. Recent studies of the QSO absorption
lines provide a 4.1 standard deviation support for this prediction.
Our considerations suggest that in the presence of
space-time fluctuations, the
principle of local gauge invariance, and the related notion of 
parallel transport, must undergo fundamental changes.
\end{abstract}


\newpage
\section{Introduction}
\label{sect:intro}

The idea put forward by St\"uckelberg and Feynman that antimatter is 
nothing more than matter propagating backward in time acquires a special 
significance in cosmology. In the absence of gravity, space-time is
a kinematic arena. On the one side, 
it constrains the form of the laws of nature by 
preserving their form under Poincar\'e transformations. On the other side, 
it determines the equations of motion and introduces 
the notion of matter and antimatter
in terms of particular finite dimensional representations of the 
Lorentz group.

As soon as one begins to take effects 
of gravitation into account, and allows for an interplay of the quantum
and gravitational realms, space-time becomes a dynamical entity.
In particular, in the early universe this entity suffered violent 
quantum-induced fluctuations as was first written down at length 
by Wheeler \cite{jw}. Once the temporal component of this primordial 
space-time foam, as the Planck-era space-time with quantum-induced 
fluctuations has come to be known, is allowed to be endowed with fluctuations 
in its direction, the question immediately arises as to what 
implications does St\"ukelberg's and Feynman's framework hold for 
the early universe.   

The evidence in favor of the cosmic matter-antimatter asymmetry has been 
strengthened by a series of recent observations. For instance, 
the antihelium-helium ratio, as reported by the 
Alpha Magnetic Spectrometer collaboration \cite{ams}, is 
\beq
\frac{\Phi_{\overline{\mbox{He}}}}{\Phi_{\mbox{He}}} <  1.7 \times 10^{-6}.
\eeq
The AMS detected no antihelium, or any $Z\ge 2$ anti-nuclei (which could 
have been created in an antimatter star). When this information is
coupled with the observed isotropy of the 
cosmic microwave background, on the one hand,
and the information on the cosmic diffuse 
$\gamma$-ray background, on the other hand, one comes to the 
conclusion that we live in a universe
which contains no significant amount of antimatter \cite{ks}. 
This asymmetry is known as the cosmic antimatter problem, or
problem of the cosmological CP violation.
The latter interpretation implicitly assumes CPT symmetry to hold in the early
universe. However, as long as the early universe contains 
an inherent element of 
non-locality, CPT might have suffered a substantial violation \cite{grf1994}.
{}For this reason, it is perhaps not quite  appropiate to refer to the 
cosmological matter-antimatter asymmetry as the cosmological CP violation.
Stated in the texbook language, the cosmological  
matter--antimatter problem reads \cite{p}:

\begin{quote}
At $kT \gsim m_p c^2$, there exist in equilibrium roughly equal 
numbers of photons, protons and antiprotons. Today, $N_p/N_\gamma
\sim 10^{-9}$, but $N_{\overline{p}}\simeq 0$. Conservation of 
baryon number would imply that $N_p/N_{\overline{p}}= 1+O(10^{-9})$
at early times. Where did this initial asymmetry come from?
\end{quote}

Our answer, in this essay, is: \textit{This initial asymmetry
is a signature of the violent space-time fluctuations in the 
early universe.}

While referring to space-time fluctuations one needs to clearly identify
the gravitational environment one is embedded in. For the present epoch, there
are at least four important studies which explore the observability of the 
space-time foam effects. In Ref.~\cite{gac1} observability of the spatial 
component of the space-time foam is studied. For certain models, the 
space-time foam has been found to carry detectable effects 
in gravity-wave interferometers. In Ref.~\cite{gac1998}, 
Amelino-Camelia \textit{et al.}
also explore the possibility of studying quantum-gravity effects via analysing
the fine-scale structure and hard spectra of gamma-ray bursters. 
The work of Klapdor-Kleingrothaus \textit{et al.}
\cite{kps} investigates  the effects of space-time foam on the neutrino-less
double $\beta$ decay and neutrino oscillations  within the astrophysical 
context. Finally, a series of papers \cite{a,b,c} have attempted to understand
the existing data  on solar and atmospheric neutrinos in terms of 
quantum-gravity de-coherence induced by space-time fluctuations.

LoSecco has analysed the SN1987a ``neutrino''  events observed in
the Kamiokande and Irvine-Michigan-Brookhaven (IMB) detectors.
Assuming that these events contain $\nu_e e$ scattering events
(i.e., $\nu_e e \rightarrow \nu_e e$), as well as $\overline\nu_e p$
capture events (i.e., $\overline\nu_e p \rightarrow e^+ n$), he verified
CP invariance in general relativity to a few parts in $10^6$
[see, Ref. \cite{LoSecco} for exact meaning of this number]. However, 
it is to be noted that the LoSecco limit on CP violation in general 
relativity applies to the present epoch only, i.e., when the space-time
fluctuations are ``tiny.'' It does not hold
for the early universe where one expects violent fluctuations of space-time.
It is that very epoch of the universe that is of prime interest to us here.

To establish
the thesis that the observed cosmological matter-antimatter asymmetry
arises, and carries a significant contribution from  
asymmetric space-time fluctuations and their interplay with the 
St\"uckelberg-Feynman interpretation of antimatter, it is essential 
to first undertake an \textit{ab initio} look at the fermionic 
(baryonic/leptonic) representation space from the point of view of the 
Lorentz symmetry without space-time fluctuations. This is done in 
Section~\ref{Sect:B}. Therein we shall discover a new quantum-mechanical
phase in Sec.~\ref{Sect:B1}. 
In Section~\ref{Sect:B3} we shall connect this phase 
with the matter-antimatter symmetry. The stated answer is then established in
Section~\ref{Sect:C} by invoking the St\"uckelberg-Feynman 
interpretation of antimatter \cite{s,f}.  
Section~\ref{Sect:C2} briefly studies the effect of the space-time fluctuations
on the fine structure constant and concludes that it must have been smaller
in the past. Encouragingly, recent studies of the QSO absorption lines 
provide a 4.1 standard deviation support for this prediction.
The essay closes by a few concluding remarks in 
Section~\ref{Sect:D}.

\section{Spin-${\bf 1/2}$ matter-antimatter without space-time fluctuations}
\label{Sect:B}

In order to study implications of space-time fluctuations on the 
notion of matter and antimatter we here undertake a critical and detailed 
look at the $(1/2,0)\oplus(0,1/2)$ representation space.

\subsection{Matter-antimatter phase factor}
\label{Sect:B1}

\def\vp{\vec p\,}
\def\v0{\vec 0\,}
\def\beq{\begin{eqnarray}}
\def\eeq{\end{eqnarray}}

The spin-$1/2$ fermions of interest are described by the 
$(1/2,0)\oplus(0,1/2)$ representation space of the 
Lorentz group (in the usual notation of Ref.~\cite{lhr}). 
This assumption shall be taken to provide sufficiently
accurate description of baryons and leptons even in the early universe  ---
its validity may, at most, be confined to sufficiently small local domains. 
In general, however, an important distinction
must be made whether we consider the  $(1/2,0)\oplus(0,1/2)$ representation
space of the Dirac type, or that of Majorana type. Fortunately,
the arguments that we present are equally valid for the Dirac case as for the
Majorana case. Because some of the readers may be more 
familiar with the usual Dirac construct, 
and because certain additional subtleties enter the Majorana construct 
\cite{dva}, we shall confine our attention to the 
$(1/2,0)\oplus(0,1/2)$ Dirac construct.

To describe  charged particles
in the Dirac sense the appropriate $(1/2,0)\oplus(0,1/2)$ 
spinors have the form:
\beq
\psi(\vp) \equiv \left(
			\begin{array}{c}
			\phi_R(\vp)\\
                        \phi_L(\vp)
                        \end{array}
                   \right).\label{ds}
\eeq
The $\phi_R(\vp)$ transforms as a $(1/2,0)$ spinor,\footnote{
Note that identical transformation properties under Lorentz group
do not necessarily imply that other transformations properties
will be identical as well. The latter, e.g., may refer to transformations
under C, P, and T. Thus, the Dirac- and Majorana-$(1/2,0)\oplus(0,1/2)$
constructs carry different physical characteristics under operations
of C, P, and T, while carrying identical transformations 
under the Lorentz group.}
and boosts as 

\beq
\phi_R(\vp) = \exp\left(\frac{\vec\sigma}{2}\cdot\vec\varphi\right)
\phi_R(\v0).\label{r}
\eeq
In the above equation, the $\vec \sigma$ stand for the standard Pauli 
matrices, and $\vec\varphi$ is the boost parameter. 
The $\vec 0$ corresponds to the momentum  
of a  particle at rest.
The definition of $\vec\varphi$
is motivated by the fact that $E^2-
\vp^2 = m^2$, and that $\cosh^2\alpha -\sinh^2\alpha=1$ (as an
identity). Thus, one can parameterize Lorentz boosts via  $\vec\varphi$,
\beq
\cosh\varphi=\frac{E}{m}, \quad \sinh\varphi=\frac{\vert\vp\vert}{m},\quad
\widehat{\varphi} = \frac{\vec p}{\vert\vp\vert}.\label{bp}
\eeq
We shall assume that $m\ne 0$.
On the other hand,  $\phi_L(\vp)$ transforms as a $(0,1/2)$ spinor, 
and boosts with the opposite sign in the exponent:  
\beq
\phi_L(\vp) = \exp\left(- \frac{\vec\sigma}{2}\cdot{\vec\varphi}\right)
\phi_L(\v0).\label{l}
\eeq
As we shall see in Sec.~\ref{Sect:B2},
the wave equation satisfied by the $(1/2,0)\oplus(0,1/2)$ 
spinor defined in Eq. (\ref{ds}) follows from: 
(a) The transformation properties of $\phi_R(\vp)$ and $\phi_L(\vp)$ 
as contained in Eqs. (\ref{r}) and (\ref{l}), and
very importantly, (b) The relative phase between the  $\phi_R(\v0)$   and 
 $\phi_L(\v0)$. 
The latter observation  is of special importance to us here
and it shall be found to lie at the heart of our arguments.

Due to the isotropy of the zero-momentum vector,
$\vec 0$, one may argue that
$\phi_R(\v0) = \phi_L(\v0)$. In fact, that is precisely what is
done in the standard textbooks \cite{lhr,H}. 
Ryder's classic book on the theory of quantum fields,
in fact, argues:
\begin{quote}
Now when a particle is at rest, one cannot define 
its spin as either left- or right-handed, so 
\beq
\label{eq:p44}
\phi_R(\v0) = \phi_L(\v0).
\eeq 
\end{quote}
Same assumption has been adopted by  Hladik \cite{H},
and we have encountered it in literature elsewhere as well (for which 
we have not kept a complete record).
The assertion (\ref{eq:p44}) is justified if one was to
confine to a purely classical framework.
However, if one is
to invoke a quantum framework for the interpretation of these spinors
then this equality can only be claimed up to a phase:
\beq
\phi_R(\v0) = \zeta \phi_L(\v0)\, .\label{rl}
\eeq
We find that a convenient choice  for  $\zeta$ is:
\beq
\zeta= \pm \exp(\pm i\phi)\, .
\eeq
The  $+$ sign is to be taken for ``particle'' spinors, while  the $-$ sign is
to be used for the ``antiparticle'' spinors (see, Ref.~\cite{dva_nl}).
Otherwise, i.e., if one ignores this phase, one misses  
anti-particles!\footnote{In order to place this observation within a proper context, we wish to
stress that this phase becomes manifest only within the group-theoretical 
derivations of the Dirac equation. Further, in the relevant limit 
it is in agreement with the remarks of Gaioli and
Garcia Alvarez \cite{ga} on Ryder's derivation of the Dirac equation.
Moreover, if C, P, and T covariances are invoked,  
it is implicitly contained in the more traditional
treatments of the Dirac equation. However, to the best of our knowledge,
nowhere in literature has the full physical content of this phase
been understood; nor has it been 
explored in the context of space-time fluctuations. The first hint for
the possible existence of this phase appears in a footnote 
of  a 1993 paper on the 
Bargmann-Wightman-Wigner type quantum field theory  \cite{bww}.}

The most general form of $\phi$ is a $2\times 2$ space-time dependent
matrix. That is,
\beq
 \exp(i\phi) = \exp\left(i\sigma^a \phi_a(t,\vec x)\right)
\eeq
where $\sigma^a$ forms the set $(I_2,\vec\sigma)$ with $I_2$ as
a $2\times 2$ identity matrix, and the $\phi_a(t,\vec x)$ 
are a set of four space-time dependent, i.e., local, parameters.
The demand that the resulting spinor of Eq.~(\ref{ds}) still transform
as an $(1/2,0)\oplus(0,1/2)$ object yields, $\phi_a=0$, for $a=1,2,3$.
Consequently, $\phi$ is reduced to an identity matrix multiplied by a 
space-time dependent real parameter $\phi_0$:
\beq
\phi = I_2\times\phi_0(t,\vec x)
\eeq 
In the absence of space-time fluctuations, a glimpse into the
physical and  mathematical content of $\phi$ can already be 
obtained by treating $\phi_0$ as a  real c-number with no space-time
dependence. By doing so, the effect of space-time fluctuations --- 
a subject of our immediate interest --- shall become easier to 
understand.

\subsection{Derivation of a Dirac-type wave equation}
\label{Sect:B2}

Equations (\ref{r}), (\ref{l}), and (\ref{rl})  contain 
essentially
the entire kinematic structure 
of the \textit{Dirac-type} spin-$1/2$ particles. To see
this we follow the foot steps of Lewis Ryder \cite{lhr}, 
but we now carefully incorporate the important ingredient embedded in a 
non-vanishing $\phi$.  


\begin{enumerate}
\item

On the right-hand side of Eq. (\ref{r}), substitute for $\phi_R(\v0)$
from (\ref{rl}). This gives
\beq
\phi_R(\vp)=\zeta \exp\left(\frac{\vec\sigma}{2}
\cdot\vec\varphi\right)
\phi_L(\v0). \label{a}
\eeq

\item
From Eq. (\ref{l}) obtain, 
\beq
\phi_L(\v0)=
\exp\left(\frac{\vec\sigma}{2}\cdot\vec\varphi\right)\phi_L(\vp),
\eeq
and insert it into the right-hand side of 
Eq. (\ref{a}). This yields:
\beq
\phi_R(\vp)=\zeta\exp\left(\vec\sigma\cdot\vec\varphi\right)
\phi_L(\vp).\label{b}
\eeq

\item
Similarly, starting from Eqs. (\ref{l}) and (\ref{rl}) we obtain:
 \beq
\phi_L(\vp)=\zeta^{-1}\exp\left(-\vec\sigma\cdot\vec\varphi\right)
\phi_R(\vp).\label{c}
\eeq

\item
Now, because $
\left(\vec\sigma\cdot\widehat{p}
\right)^2 = 2\times 2\,\,\mbox{Identity
matrix,}\,\, I_2$
\beq
\left(\vec\sigma\cdot\widehat{p}
\right)^n=\cases{I_2 & for n even\cr
                             \vec\sigma\cdot\widehat{p} 
& for n odd\cr}.
 \label{Pauli}
\eeq
This leads to the identities:\footnote{Note, it is precisely the 
property expressed by Eq.~(\ref{Pauli}) that is responsible for the
\textit{linearity} of the Dirac equation as will become obvious below.}

\beq
\exp\left(\pm \vec\sigma\cdot\vec\varphi\right)
=\frac{E I_2\pm\vec\sigma\cdot\vec p}{m}
\eeq

\item
Next, substitute these identities in Eqs. (\ref{b}) and (\ref{c}), 
and re-arrange to obtain:
\beq
\left(
\begin{array}{cc}
-m \zeta^{-1} & E I_2 +  \vec\sigma\cdot\vec p \\
E I_2- \vec\sigma\cdot\vec p & -m \zeta
\end{array}
\right)
\left(
\begin{array}{c}
\phi_R(\vp)\\
\phi_L(\vp)
\end{array}
\right)= 0.
\eeq
\item
Finally, with $p_\mu=(p^0, \,-\vp)$, $E=p^0$, 
read off the Weyl-representation
gamma matrices:\footnote{We now abbreviate 
$I_2$ by $I$. The zeros below stand for
$2\times 2$ null matrices, and $\sigma^1=\sigma_x$, etc.}
\beq
\gamma^0= 
\left(
\begin{array}{cc}
0 & I\\
I & 0
\end{array}
\right), \quad 
\gamma^i= 
\left(
\begin{array}{cc}
0 & - \sigma^i\\
\sigma^i & 0
\end{array}
\right),
\eeq
and introduce
\beq
\Phi(\phi)= 
\left(
\begin{array}{cc}
\zeta^{-1}(\phi) & 0\\
0 & \zeta(\phi)
\end{array}
\right)\, .
\eeq
This yields,
\beq
\Big(\gamma^\mu p_\mu - m\Phi(\phi)\Big)\psi(\vec p) =0.
\label{eq}
\eeq

\end{enumerate}


The obtained equation is Poincar\'e covariant and indeed carries
the solutions with the correct dispersion relations $E=\pm\sqrt{\vp^2+m^2}$,
because  
\beq
\mbox{Det}\left[\gamma^\mu p_\mu - m\Phi(\zeta)\right]
=\left(\vp^2+m^2-E^2\right)^2.
\eeq
Now, if $\Phi(\zeta)$ is allowed to carry space-time dependence, the same 
method proceeds (with $\phi_0$ becoming a local parameter), and 
Eq. (\ref{eq}) becomes the CP-violating Dirac equation \textit{postulated}
in Ref. \cite{funakubo}.

\section{New phase and matter-antimatter symmetry}
\label{Sect:B3}

The reader may have already noted that
$\mbox{Det}\left[\gamma^\mu p_\mu - m\Phi(\phi)\right]$
is independent of $\phi$. Consequently,
Poincar\'e covariance alone
cannot constrain $\phi$. To constrain this phase angle 
one needs to invoke additional symmetries.
If we demand Eq. (\ref{eq}) to be covariant under the operations of C, P, and T, 
then we find that 
\beq
\phi_0= n \times 2\pi,\quad \mbox{with}\,\, n=0,1,2,\ldots.\label{n}
\eeq
That is, for a kinematic structure that is covariant under C, P, 
and T symmetries, Eq. (\ref{rl}) reduces to
\beq
\phi_R(\v0) = \pm \phi_L(\v0). \label{rl2}
\eeq
The $\psi(\vp)$ of Eq. (\ref{ds}) constructed with a plus sign in the above
equation results in the ``particle $u$-spinors'', while the minus sign 
in Eq. (\ref{rl2}) results in the ``antiparticle $v$-spinors.'' Explicitly, the 
particle spinors are
\beq
u_{+1/2}(\vp) = \kappa(\vp)
\left(
\begin{array}{c}
1\\
0\\
1\\
0
\end{array}
\right),\quad
u_{-1/2}(\vp) = \kappa(\vp)
\left(
\begin{array}{c}
0\\
1\\
0\\
1
\end{array}
\right)
\eeq
and the antiparticle spinors read
\beq
v_{+1/2}(\vp) = \kappa(\vp)
\left(
\begin{array}{c}
1\\
0\\
-1\\
0
\end{array}
\right),\quad
v_{-1/2}(\vp) = \kappa(\vp)
\left(
\begin{array}{c}
0\\
1\\
0\\
-1
\end{array}
\right)\, . \eeq
Here the $(1/2,0)\oplus(0,1/2)$ boost, $\kappa(\vp)$, as contained 
in Eqs. (\ref{r}) and (\ref{l}), is given by
\beq
\kappa(\vp)=
\left(
\begin{array}{cc}
\exp(\vec\sigma\cdot\vec\varphi/2) & 0 \\
0 & \exp(-\vec\sigma\cdot\vec\varphi/2)
\end{array}
\right).
\eeq
The particle-antiparticle spinors found in the standard textbooks, see, e.g.,
Bjorken and Drell's (BD) well-known classic \cite{bd}, 
are related to those obtained here by:

\beq
\psi^{BD}(\vp )=\frac{1}{\sqrt{2}}
\left(
\begin{array}{cc}
I & I \\
I & -I
\end{array}
\right)\,\psi (\vp)
\eeq

Parenthetically, one may be interested to note that different 
fermions in Nature, i.e., quarks and leptons, need not carry same 
$\phi_0$ dependence for the underlying spinorial structure. In the event, 
say, the $s$-quark carried a  different $\phi_0$ than given by 
Eq. (\ref{n}),
the resulting kinematic structure might not be intrinsically CP respecting.
We envisage the above considerations, when incorporated in the kinematic 
structure of the standard model particles, to hold serious potential to provide
a source for CP violation in particle physics. But, here in this essay 
our interest is mainly in the cosmological matter-antimatter asymmetry ---
a violation which we now argue arises from an interplay of the $\phi_0$
and space-time fluctuations.

At this stage in our essay 
we have arrived at  the standard $(1/2,0)\oplus(0,1/2)$
Dirac construct. The merit of this apparently simple exercise 
lies in having exposed a CPT-related hidden phase in the  
$(1/2,0)\oplus(0,1/2)$
representation space and having deciphered that  
the Poincar\'e symmetry alone does not entirely constrain
the $(1/2,0)\oplus(0,1/2)$ representation space. One needs additional
requirements of P, C, and T symmetries. 
Further, invoking the standard
conventions we reach the key understanding. It reads:

\begin{quote}
Spin-1/2 matter of the Dirac type corresponds to the \textit{plus}
sign in Eq.~(\ref{rl2}), while the antimatter is characterized
by the \textit{minus} sign. 
\end{quote}

Now, in order to arrive at our thesis, we make the needed 
observations explicitly. In the standard textbook treatments, 
the Dirac's hole theory and 
St\"uckelberg-Feynman's interpretation of antiparticles as particles
going backward in time are considered equivalent. However, as 
pointed out by Hatfield \cite{bh}, and confirmed explicitly by
the work on $(1,0)\oplus(0,1)$ representation space \cite{bww},
the former framework applies only to the fermions while the
latter is applicable to fermions as well as to bosons. 
From a parenthetic remark of Feynman contained in his paper
entitled ``The theory of positrons'' it is also apparent that he was
aware of this advantage of his theory (see, p. 750, in \cite{f}).
For this
reason, we shall here adopt the St\"uckelberg-Feynman framework
for the matter-antimatter solutions of the  
$(1/2,0)\oplus(0,1/2)$ representation space. 
It is then seen that the $+\exp(+i\phi)=+1$ corresponds to a
particle propagating in the \textit{forward} direction of time
while the  $-\exp(-i\phi)=-1$ corresponds to the particle
propagating \textit{backward} in time (which is then interpreted as
antiparticle).

\section{Matter-antimatter asymmetry and Space-time fluctuations }
\label{Sect:C}

In the presence of space-time fluctuations, 
St\"ckelberg-Feynman framework suggests a natural extension. We 
propose this to be:
\begin{quote}
A particle moving forward in time on encountering a region of
space-time, with a backward temporal fluctuation, becomes an antiparticle.
\end{quote}
That is, temporal fluctuations 
interchange $(1/2,0)\oplus(0,1/2)$ sectors associated with the phases 
$+\exp(+i\phi)$ and $-\exp(-i\phi)$, and thereby transforms a particle into
an antiparticle:
\beq
&&\mbox{\sc Temporal fluctuations}: \nonumber\\
&&(1/2,0)\oplus(0,1/2)\,\, \mbox{Sector with}\,\, 
+\exp(+i\phi) \nonumber\\
&&\qquad\qquad\leftrightarrow 
(1/2,0)\oplus(0,1/2)\,\, \mbox{Sector with}\,\,
-\exp(-i\phi)
\eeq
In this process, the conservation of charges connected by the C operator
(such as, electric charge and baryon number) are  
violated.
With CP symmetry and baryon number conservation violated in this manner these
processes must proceed in an environment with a  slight lack of 
thermal equilibrium in order to account for the cosmological 
matter-antimatter \cite{Sakharov} asymmetry. 
We underline ``slight'' in order to 
accommodate for the high degree of temperature isotropy of the 
cosmic microwave background.

In the cosmological realm the advantage of the St\"uckelberg-Feynman
over the Dirac's hole theory is more than formal. In fact, the the former
framework appears crucial to understanding the matter-antimatter 
asymmetry in the early universe. To argue this assertion we note:
\begin{enumerate}
\item
In the standard big bang scenario, at $t=0$, it is not possible to define
propagation backward in time. 

\item
The interplay of the quantum and gravitational realms implies that space-time
itself must carry fluctuations. The simplest argument leading to this 
observation is that if gravitational 
effects associated with a quantum measurement
process are incorporated, then space-time measurements become non-commutative.
This has been discussed in detail, e.g., in 
Refs.~\cite{grf1994,sd,ak,fs,as,ns,pla2000,gac}.
\end{enumerate} 
Now photons are self-conjugate under the operation of charge conjugation, C,
and thus are insensitive to the arrow of time. On the other hand,
in the St\"uckelberg-Feynman framework, spin-$1/2$ matter fields of the Dirac
type are endowed with a sensitivity to the arrow of time.  
To understand  the cosmological matter-antimatter asymmetry,
the observation in item (1) above suggests to confine one's attention
to a universe with only fermions (and no anti-fermions) at $t=0$.
This is precisely what we shall do next. 
The possibility of a photonic birth shall be taken up in Sec. \ref{Sect:C1}.

Given this circumstance, and once the reality of quantum-induced space-time
fluctuations is accepted, it is readily seen that from the vantage point of
an hypothetical observer riding the temporal component of these fluctuations,
the St\"uckelberg-Feynman interpretation implies that matter-antimatter 
identity of the primordial spin-$1/2$ Dirac type particles varies with 
time.\footnote{An alternate view may be taken that an $e^-$, say [described by
a $u(\vp)$], propagating forward in time encounters a space-time region
undergoing a backward temporal fluctuation. The effect of this encounter is
to transform an $e^-$ into an $e^+$ [described by a $v(\vp)$].
It is be noted that in every space-time region there are, with respect 
to the local arrow of time, both forward and backward in time propagating 
particles.} That is, as seen by the indicated hypothetical observer, 
the phases in Eq. (\ref{rl2}) is no longer uniquely plus, or uniquely minus,
for a given particle. But, instead, it oscillates, $+\leftrightarrow -$, 
with time 
in an observer-dependent manner. As determined by the averaged observations 
of a large number of indicated hypothetical 
observers\footnote{Whose ride on the space-time fluctuations is 
assumed incoherent.} the ratio defining the cosmic antimatter problem
\beq 
\frac{N_p}
{N_{\overline{p}}}
{\Bigg\vert}_
{kT \approx m_p c^2}
= 1+O(10^{-9})    
\eeq
is then an indication that the temporal fluctuations
were dramatically violent even at $kT \approx m_p c^2$. The early universe
at $kT \approx m_p c^2$ was characterized by a forward-backward
asymmetry in time of the order $O(10^{-9})$. This asymmetry was perhaps
even much smaller at still higher temperatures. The 
cosmological arrow of time was still in the birth pangs.
However, it is apparent that as soon as the cosmological arrow of time becomes
well pronounced, i.e., the backward temporal fluctuations loose their 
amplitude in relation to the forward fluctuations, the universe appears as 
composed of matter.

In brief, the considered scenario, the primordial universe begins with a 
maximal matter-antimatter asymmetry in favor of matter. 
However, the violent space-time fluctuations soon transform 
it into a matter-antimatter
universe with roughly equal densities of both. At $t\approx m_p\, c^2$, 
the matter-antimatter asymmetry is roughly one part in a billion. 
At this stage, matter and antimatter annihilate and in the process an attempt towards 
a thermal equilibrium of various then-existing radiation and particle components can proceed. 
To what extent a thermal equilibrium is reached is determined by $t_u$ (see Eq. (\ref{tu}) below), 
and various related thermodynamic considerations.

Adding to thermal equilibrium is the fact that before a distinct cosmological arrow 
of time is born, cosmic time is really not a good measure of the cosmic age. To attend
to this unique cosmolgical circumstance, we suggest that the cosmic age
should be defined as the sum of the cosmic time plus a \textit{thermodynamic time}. Around
$t=0$, we define the thermodynamic time,  
$t_{th}$, as
\beq
t_{th}\approx \bar A \times N(t)\, .
\label{th}
\eeq 
Here, $\bar A$ stands for the
average amplitude of the temporal fluctuations (around $t=0$), while
$N(t)$ equals the number of the accummulated fluctuations at cosmic time $t$.
{}From that, the age of the universe $t_u $ can be read off as
\beq
t_u \approx t+t_{th}\, .
\label{tu}
\eeq
In order that a cosmological arrow of time comes into a distinct existence
we surmise that (and a full theory of quantum gravity should account for)
as evolution of the universe proceeds, the amplitude associated with the
backward temporal fluctuation begins to diminish compared to the amplitude
for forward fluctuations. As this happens, the universe once again begins to
appear as matter dominated, but now it has an additional radiation component.
$N(t)$ can, in principle, 
be determined from the constraints imposed by directional
uniformity of the temperature of the cosmic microwave background radiation, and
additional sources of thermalization that may emerge from ``opening up'' of the
light cone in the theory of relativity that carries an invariant length 
\cite{gac_lc}.

\subsection{Cosmological matter-antimatter asymmetry with a photonic birth}
\label{Sect:C1}

An alternate scenario would be a photonic birth of the universe, followed
by a pair-created matter and antimatter. As a result of the asymmetric 
temporal fluctuations, the matter-antimatter ratio will undergo local changes
resulting into excess of one type of matter over the other. This can happen
when, e.g., a $e^+$ encounters a region of space-time where a temporal 
fluctuation has changed the direction of time.  
In that process the $e^+$ becomes an $e^-$, which can thus, in one of the 
possibility (say), annihilate an $e^+$ which has either
{\textit not suffered any reversal} in temporal fluctuation, or,
has suffered an {\textit even number} of them. 
In considering such a process one may envisage 
a rigid mountain form placed on two spatial dimensions and one time 
dimension  --- the latter representing the height. 
In such processes 
space-time fluctuations result in a net creation/destruction of electric 
charge (and/or, baron number). Note that a similar fate meets if we begin
with an $e^-$ in the above example. A one part in a billion imbalance in such 
processes could then give rise to the observed matter-antimatter asymmetry.
This scenario is conceptually more intricate and calls for a 
detailed Monte Carlo simulations to explore its viability.

\subsection{Fine Structure constant and space-time fluctuations}
\label{Sect:C2}

Above considerations suggest a natural and testable prediction: 

As already noted, 
it is a general expectation of every model of space-time
foam that space-time fluctuations were much more violent
at the big bang, and that these have slowly become less
intense. As a consequence, far from big bang, and
still in the distant past (to become more precise below), 
the magnitude of the 
effective charge of an electron would appear smaller 
than its present value. For,
while the time fluctuates backward, a particle that was interpreted 
as an electron would appear as a positron. 
Larger backward time fluctuations in the past would thus
effectively reduce the magnitude of the observed electronic charge. 
The same result applies to a proton. Thus, under the assumption
of time-constancy for $\hbar$ and
$c$  over the period of interest, one would expect 
the fine structure constant $\alpha=e^2/\hbar c$ to be smaller in
the past. 

As this essay was written, we learned that this prediction is consistent 
with the latest results on the variation of the fine structure 
constant \cite{alpha}. With the definition,
$
\Delta \alpha/\alpha = (\alpha_z-\alpha_0)/\alpha_0$, where $\alpha_0$
is the present day value of $\alpha$ and $\alpha_z$ is the value at absorption
redshift $z$ [see \cite{alpha} for details], they find:
\beq
\label{eq:neg_alpha}
\frac{\Delta\alpha}{\alpha} = (-0.72\pm 0.18) \times 10^{-5}
\eeq 
over the redshift range $0.5 < z < 3.5$. The result (\ref{eq:neg_alpha})
confirms their earlier study \cite{earlier,w99}, 
and now represents a statistically significant 
$4.1$ standard deviation effect in favor of the conclusion that
$\alpha$ was slightly smaller in the past.

\section{Concluding remarks}

\label{Sect:D}
The notion of charge conservation, and in particular that of baryon
number conservation, is an empirically derived concept from observations
in an cosmic epoch when the cosmological arrow of time is a 
well established object. However, as we have seen, these conservations law
carry little meaning when this very important object was still in its
formative stages and suffered violent fluctuations. 
This circumstance suggests that, in the presence of
space-time fluctuations, the
principle of local gauge invariance (whether in some internal space, or
space-time), and the related notion of 
parallel transport, must undergo fundamental changes.

Given a specific
form of the space-time fluctuations (an input to come from a 
future full theory of quantum and gravitational realms), one may derive the 
time-dependent matter-antimatter ratio beginning with, 
say, matter alone. The considerations presented above strongly indicate
that even at  $kT \approx m_p c^2$ the space-time fluctuations 
were so violent that for every observer the matter-antimatter ratio 
appeared as to be roughly unity. 
As the backward fluctuations in time became less dominant,
and the cosmic arrow of time acquired a physical reality, the same 
universe began to appear as matter dominated. The violent 
interplay of the quantum and gravitational realms did not respect 
conservation of electric charge, and baryon number. Interestingly, 
all this is a natural consequence if one adopts the  
St\"uckelberg-Feynman interpretation of matter-antimatter and allows
for quantum-induced fluctuations of space-time. If all this
provides a fundamental space-time origin of the
cosmological matter-antimatter asymmetry, 
the CP violation in heavy quark systems 
may be understood if the the light quarks carried a $\phi_0$ given by
Eq. (\ref{n}),
while the heavy quarks, say the strange one, carried a 
CP violating kinematic structure.

In the last two years it has become apparent that 
quantum-gravity induced fluctuations in \textit{spatial} distances
can be studied using gravity-wave interferometers \cite{gac1,gac2,gac3,jn}.
This essay indicates that a careful analysis of the
cosmological evolution of the fine structure constant, along the lines
of the pioneering work of   
Webb {\em et al.}~\cite{alpha}, could become a powerful 
probe of the \textit{temporal} fluctuations in the space-time foam.

\section*{Acknowledgments}
Giovanni Amelino-Camelia, Naresh Dadhich, Lorenzo Diaz,
Gaetano Lambiase, Yong Liu, J. M. LoSecco, and P. Singh
provided comments on an earlier version of this essay.
We extend to them our thanks.

\end{document}